\documentclass[aps,prb,twocolumn,superscriptaddress,10pt,article,nofootinbib,showpacs,longbibliography,nofootinbib]{revtex4-2}
\usepackage{blindtext}
\usepackage{lipsum}
\usepackage{graphics}
\usepackage{amsmath}
\usepackage{graphicx}
\usepackage{graphics}
\usepackage{amssymb}
\usepackage{verbatim}
\usepackage{physics}
\usepackage{float}
\usepackage[normalem]{ulem}
\usepackage[colorlinks, citecolor=blue]{hyperref}
\usepackage[dvipsnames]{xcolor}
\usepackage{bbm}
\usepackage{enumitem}   

\begin{document}

\title{Comment on ``Traversable wormhole dynamics on a quantum processor''\vspace{-2mm}}

\author{Bryce Kobrin}
\affiliation{Department of Physics, University of California, Berkeley, CA 94720, USA}

\author{Thomas Schuster}
\affiliation{Department of Physics, University of California, Berkeley, CA 94720, USA}

\author{Norman Y. Yao}
\affiliation{Department of Physics, University of California, Berkeley, CA 94720, USA}
\affiliation{Department of Physics, Harvard University, MA 02138, USA}

\begin{abstract}
A recent article [Nature {\bf 612}, 51–55 (2022)]~\cite{jafferis2022traversable} claims to observe traversable wormhole dynamics in an experiment.
This claim is based upon performing a teleportation protocol using a Hamiltonian that consists of seven Majorana fermions with five fully-commuting terms.
The Hamiltonian is generated via a machine-learning procedure designed to replicate the teleportation behavior of the Sachdev-Ye-Kitaev (SYK) model.
The authors claim that the learned Hamiltonian reproduces  gravitational dynamics of the SYK model and demonstrates gravitational teleportation through an emergent wormhole.
We find: (i) in contrast to these claims, the learned Hamiltonian does not exhibit thermalization; (ii) the teleportation signal only resembles the SYK model for operators that were used in the machine-learning training; (iii) the observed perfect size winding is in fact a generic feature of small-size, fully-commuting models, and  does not appear to persist in larger-size fully-commuting models or in non-commuting models at equivalent system sizes.

\end{abstract}

\maketitle
 
The holographic principle posits that certain quantum mechanical Hamiltonians are dual to quantum theories of gravity~\cite{hooft1993dimensional,susskind1995world,maldacena1999large}. 
Recently, there has been tremendous interest towards \emph{experimentally} realizing such quantum mechanical models on a quantum processor. 
To this end, an important development was the discovery of a quantum teleportation protocol that is related to traversable wormholes (Fig.~\ref{fig:teleport}a)~\cite{gao2017traversable,maldacena2017diving,gao2021traversable,brown2019quantum,nezami2021quantum,schuster2022many}. 
Specifically, when the teleportation protocol is implemented using a  Hamiltonian that is holographically dual to gravity, successful teleportation is described from the dual perspective as a particle traveling through a  traversable wormhole.

A recent article~\cite{jafferis2022traversable} claims to observe traversable wormhole dynamics in an experimental setting.
The most direct way to observe traversable wormhole dynamics would be to experimentally implement the SYK model, which is dual to  gravity~\cite{sachdev1993gapless,kitaev2015simple}.
However, it is extremely challenging to experimentally implement even a small-size version of the SYK model.
To this end, \cite{jafferis2022traversable} uses a machine-learning procedure to construct a sparse Hamiltonian that aims to preserve gravitational physics.
More specifically, the machine-learning procedure is based upon reproducing the teleportation behavior of the SYK model (at system size $N=10$) with only a small number of Hamiltonian terms.
The result is the following Hamiltonian, henceforth ``Model 1'',
\vspace{-1mm}
\begin{align} 
H = & - 0.36 \psi^1 \psi^2 \psi^4 \psi^5 +0.19 \psi^1 \psi^3 \psi^4 \psi^7 -0.71 \psi^1 \psi^3 \psi^5 \psi^6 \nonumber \\
& + 0.22 \psi^2 \psi^3 \psi^4 \psi^6+0.49 \psi^2 \psi^3 \psi^5 \psi^7.
\label{eq3}
\end{align}
Here, $\psi^i$ are Majorana fermions satisfying $\{\psi^i,\psi^j\} = \delta_{ij}$.

The authors claim that Model 1 ``is consistent with gravitational dynamics of the dense SYK Hamiltonian'' and demonstrates ``gravitational teleportation...by means of an emergent wormhole''.
They analyze five key properties of traversable wormhole physics:
\begin{enumerate}[label=\textbf{(\roman*)},nosep]
\item scrambling and thermalization dynamics
\item a teleportation signal that is consistent with a negative energy shockwave
\item perfect size winding
\item  causal time-ordering of teleported signals
\item a Shapiro time delay
\end{enumerate}

\noindent These claims are surprising given that:
\begin{itemize}
\item[] Model 1 is fully-commuting. Each of the five Hamiltonian terms commutes with every other term. 
\end{itemize}
which is alluded to in the Supplemental Material of \cite{jafferis2022traversable}.
This property is distinct from the SYK model, and
fully-commuting models are known to exhibit markedly different dynamics from non-commuting models.

\begin{figure*}[t]
\includegraphics[width=\linewidth]{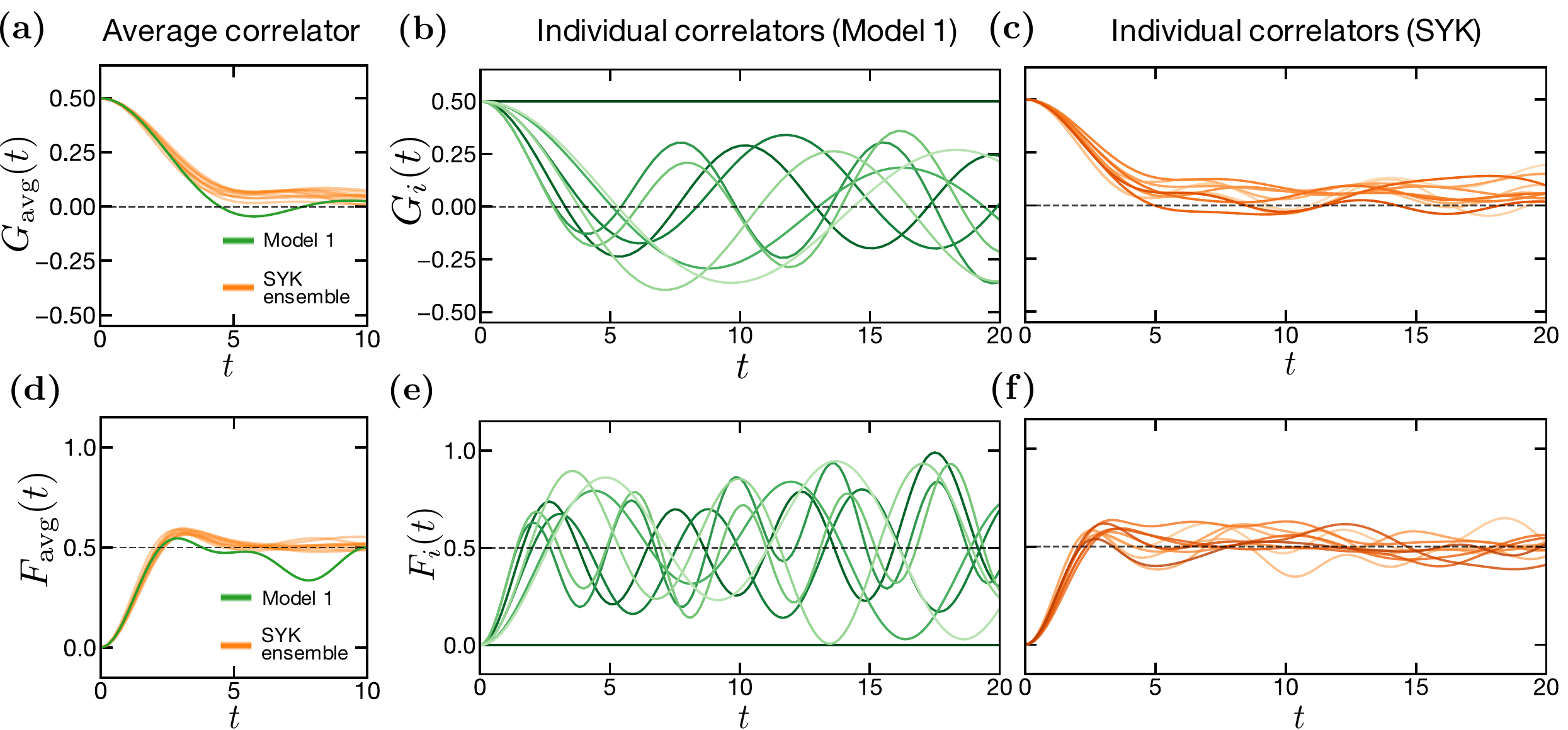}
\caption{%
\textbf{Lack of thermalization in  Model 1.}
\textbf{(a)} Two-point correlation functions averaged over Majorana operators, $G_\textrm{avg}(t)$, for Model 1 (green; replicating Fig.~3b of \cite{jafferis2022traversable}) and several disorder realizations of the $N=10$ SYK model (orange).
As observed in~\cite{jafferis2022traversable}, the average correlation function displays similar behavior between the two models.
For both models, $\beta = 4$, and, for the SYK model, the couplings are drawn from a normal distribution with mean zero and variance $6 J^2/N^3$, where $J=1.125$.
\textbf{(b)} In Model 1, the \emph{individual} two-point correlation functions, $G_i(t)$, display large oscillations.
\textbf{(c)} In the SYK model (taking a single disorder instance), the individual correlation functions all exhibit decay. This  behavior is independent of the disorder realization. 
\textbf{(d-f)} Analogous results for the average and individual four-point correlation functions, $F_\textrm{avg}(t)$ and $F_i(t)$, in Model 1 and the $N = 10$ SYK model.
Again, the agreement between the two models holds only for the averaged correlation functions \textbf{(d)}, and not for the individual correlation functions \textbf{(e-f)}.
}
\label{fig:2pt}
\end{figure*}

\begin{figure*}
\includegraphics[width=\linewidth]{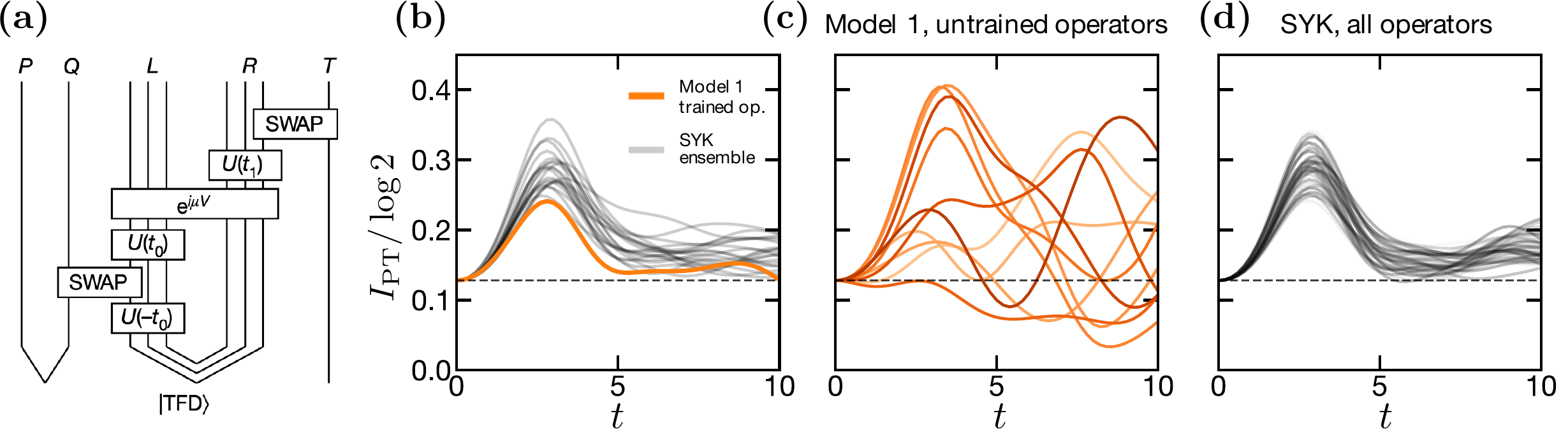}
\caption{
\textbf{Teleportation signal of Model 1.} 
\textbf{(a)} Teleportation circuit from \cite{jafferis2022traversable}.
The qubit to be teleported is swapped with a pair of Majorana operators in the left system L.
The success of teleportation from L to R is measured by the mutual information between a reference qubit P and a readout qubit T. 
\textbf{(b)} Mutual information, $I_{\textrm{PT}}$, of the  symmetric teleportation protocol with $\mu=-12$ for Model 1 (orange; replicating Fig.~2a of~\cite{jafferis2022traversable}) and several realizations of the $N=10$ SYK model with $J=1.25$ (grey). 
The machine-learning procedure in~\cite{jafferis2022traversable} trains Model 1 to reproduce the mutual information (as a function of time) of the SYK model for a specific pair of input operators, $\psi^1$ and $\psi^2$.
For this pair of operators, the mutual information  indeed shows good agreement between the two models. 
\textbf{(c)} In Model 1, when the teleportation protocol is performed with input operators that were not involved in the training procedure, i.e.~$\psi^i$ and $\psi^j$ where $i<j \in[3,7]$, the mutual information as a function of time exhibits significant variations. 
\textbf{(d)} For comparison, in the $N=10$ SYK model, the mutual information for all pairs of input operators is consistent. The dashed line indicates the mutual information at $t=0$ for reference.  
}
\label{fig:teleport}
\end{figure*}

\begin{figure*}[t]
\includegraphics[width=\textwidth]{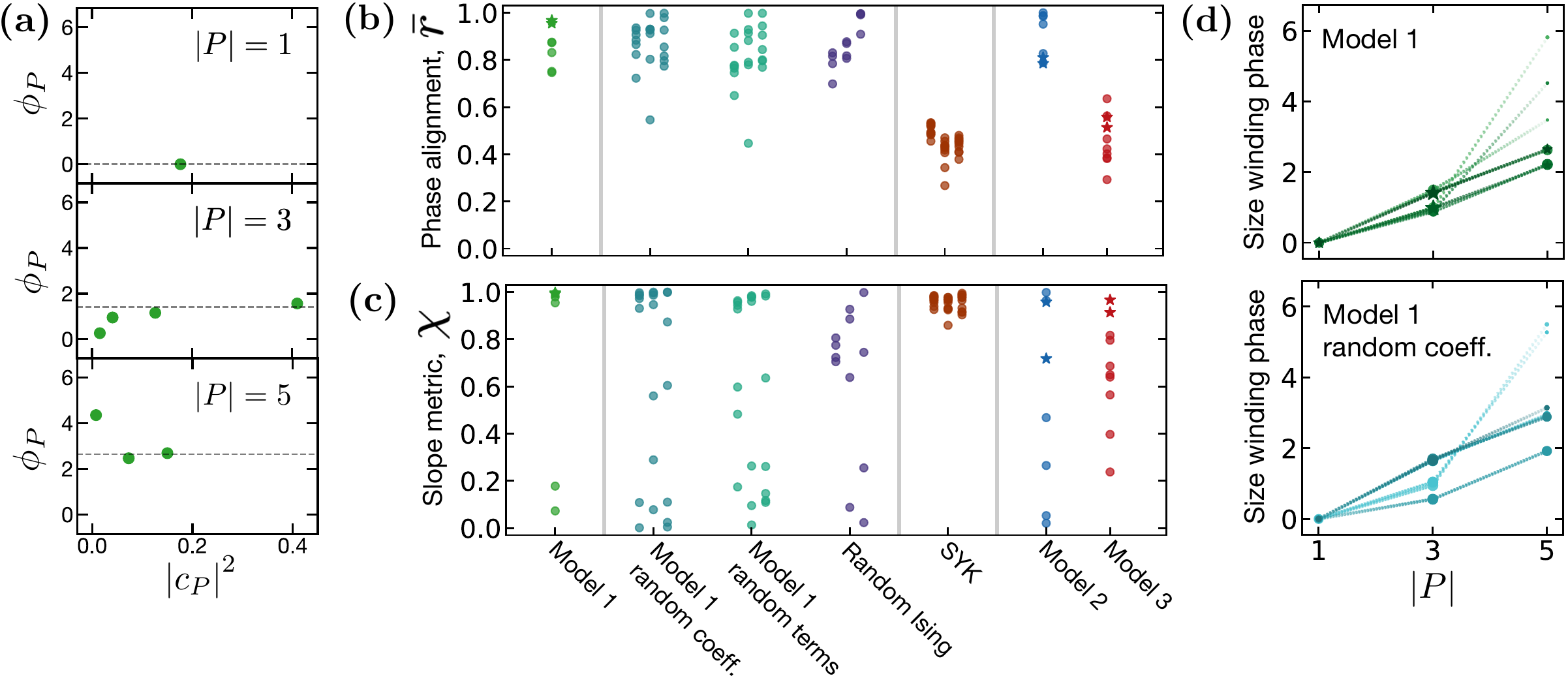}
\caption{%
\textbf{Comparison of size winding behavior in Model 1 and other random small-size fully-commuting Hamiltonians.} 
\textbf{(a)} Scatter plots depicting the eight non-zero coefficients, $c_P^2$, for $\psi^1$ of Model 1 at $t=2.8$. The x-values are the coefficient magnitudes, $|c_P|^2$, and the y-values are the coefficient phases, $\phi_P \equiv \textrm{arg }c_P^2 - \textrm{arg }q(1)$. 
Perfect phase alignment occurs when the phase of all coefficients at a given size $|P|$ matches the phase of their sum (dashed line).
This occurs trivially for the single coefficient with $|P| = 1$, and via the alignment of $\sim$2-3  coefficients for $|P| = 3,5$.
 \textbf{(b)} A comparison of the phase alignment, $\bar r$, for each operator in: Model 1, Model 1 with random coefficients, Model 1 with random terms and coefficients, a random all-to-all Ising model, the $N=10$ SYK model, Model 2, and Model 3. For Models 1,2,3, the phase alignment for the trained operators, $\psi^1$ and $\psi^2$, is indicated with a star. As in \cite{jafferis2022traversable}, we take $\beta = 4$, and time $t=2.8$ for Models 1,2 and $t=2$ for Model 3. For the random models, three different disorder realizations are shown, with small horizontal offsets for clarity.
 \textbf{(c)} An analogous comparison for the linear slope metric, $\chi$.
\textbf{(d)} The size winding phase, $\textrm{arg }q(\left | P \right |) - \textrm{arg }q(1)$, as a function of the operator size, $\left | P \right |$, for Model 1 (top) and Model 1 with random coefficients (bottom). The size of each marker is scaled proportional to $\left | q(\left | P \right |) \right |$. The stars in Model 1 correspond to operators $\psi^1$ (replicating Fig.~3d in \cite{jafferis2022traversable}) and $\psi^2$.}
\label{fig:size-winding}
\end{figure*}

Given the interest in realizing non-trivial models of quantum gravity in experiment, it seems worthwhile to investigate the extent to which the model and strategy in \cite{jafferis2022traversable} indeed capture the stated features of gravitational physics. 
Our central findings are:

\begin{itemize}[leftmargin=4.8mm]
\item In contrast to the claims of~\cite{jafferis2022traversable}, Model 1 does not thermalize. 
It exhibits strong oscillations in the  correlation functions that characterize scrambling and thermalization (Fig.~\ref{fig:2pt}). 
The  observation of thermalization in \cite{jafferis2022traversable} is  an artifact of averaging over these oscillations. 

\item  In order to generate Model 1, the machine-learning procedure in \cite{jafferis2022traversable} trains on teleportation involving two specific operators, $\psi^1$ and $\psi^2$.
The authors characterize the teleportation signal [properties {\bf (ii, iv, v)}] and size winding [property {\bf (iii)}]  only for those operators that were trained on. 
We find that the teleportation signal only resembles that of the SYK model for the specific operators that were trained on, and not for general operators that were not involved in the training (Fig.~\ref{fig:teleport}).

\item The observed perfect size winding in Model 1, is in fact, a widespread property of fully-commuting Hamiltonians at small system sizes (Fig.~\ref{fig:size-winding}). 
Putting random numerical coefficients in front of the terms in Eq.~(\ref{eq3}) or taking random commuting terms, also produces perfect size winding.
In the cases that we have examined, perfect size winding does not persist to larger system sizes or to non-commuting models at equivalent system sizes. 
 
\end{itemize}

We emphasize an inherent tension between the first and third observations: Small-size, fully-commuting Hamiltonians do not thermalize but generally exhibit perfect size winding, while the opposite is true for larger or non-commuting systems. 
None of the systems considered in \cite{jafferis2022traversable} satisfy both properties simultaneously. 
Nevertheless, both thermalization~\cite{hawking1975particle,horowitz2000quasinormal} and size winding~\cite{brown2019quantum,nezami2021quantum} are central to the holographic correspondence and are known to occur in the SYK model at large system sizes~\cite{nezami2021quantum}.

Before proceeding, we note that two additional Hamiltonians are numerically studied in the Supplemental Material of \cite{jafferis2022traversable}. We address these in Appendix~\ref{app:model2and3}.
The Hamiltonian Eq.~(S16) in \cite{jafferis2022traversable}, henceforth ``Model 2'', is produced by the same machine-learning procedure as Model 1. We find that Model 2 is nearly fully-commuting, and that the above observations similarly hold.
In contrast, the Hamiltonian Eq.~(S17) in \cite{jafferis2022traversable}, henceforth ``Model 3'', is produced by an alternate machine-learning procedure designed to maximize the difference in the teleportation signal between negative and positive couplings.
Model 3 is not fully-commuting and exhibits clearer signatures of thermalization. 
However, as noted in~\cite{jafferis2022traversable}, it does not exhibit perfect size winding.

\textbf{Thermalization and scrambling---}Ref.~\cite{jafferis2022traversable} claims that Model 1 ``scrambles and thermalizes similarly to the original SYK model as characterized by the four- and two-point correlators.''
To support this, they plot the \emph{average} of each correlator over local Majorana operators.
For example, they plot (Fig.~3b of \cite{jafferis2022traversable}) the two-point correlator $G_\textrm{avg} (t)   =  \frac{1}{8} \sum_{i=1}^8 G_i (t)$, where $G_i (t) = \textrm{Re}\left [ \left< \psi^i(t) \psi^i(0)  \right>_\beta \right ]$ and the sum is over the seven operators in Eq.~\eqref{eq3} and an additional operator $\psi^8$ that does not enter the Hamiltonian (reproduced in Fig.~\ref{fig:2pt}a). Here, $\left < \cdot \right>_\beta = \textrm{Tr}[ (\cdot) \rho_\beta]$ with $\rho_\beta = e^{-\beta H} / \textrm{Tr}[e^{-\beta H}]$.

The decay of two-point correlation functions is indicative of thermalization~\cite{dalessio2016quantum}. 
We note that it is not typical to average over a system's two-point correlation functions when exploring thermalization, since this averaging can lead to a decay that is not representative of the individual correlation functions.
As shown in Fig.~3b of \cite{jafferis2022traversable} and we reproduce in Fig.~\ref{fig:2pt}a above, for both the SYK model and Model 1, the averaged correlation function, $G_\textrm{avg} (t)$, indeed exhibits decay.
For the SYK model, this decay is consistent with the behavior of individual two-point correlation functions, and thus, thermalization (Fig.~\ref{fig:2pt}c).
However, for Model 1, the individual two-point correlators, $G_i(t)$, exhibit strong revivals as a function of time (Fig.~\ref{fig:2pt}b).
This indicates that the agreement in the thermalization behavior between the SYK model and Model 1 observed in Fig.~3b of \cite{jafferis2022traversable} is an artifact of averaging over the two-point correlation functions, and that in fact, Model 1 does not thermalize.

In Fig.~\ref{fig:2pt}d-f, we turn to the behavior of  four-point correlation functions, $F_i (t) =- \textrm{Re} \left [ \big\langle \left [\psi^i(t), \psi^i(0) \right ]^2 \big\rangle_\beta \right ]$, as well as their average, $F_\textrm{avg} (t)   = \sum_{i=1}^8 F_i (t)$.
Much like the two-point correlators, the agreement between the four-point correlation functions of Model 1 and the SYK model (Fig.~3b of \cite{jafferis2022traversable}), is an artifact of averaging. 

In the holographic correspondence, the persistent decay of the two- and four-point correlators corresponds to a perturbation falling toward a black hole~\cite{horowitz2000quasinormal,shenker2014black}. The strong revivals in the correlators of Model 1 contrast with this physics.

\textbf{Teleportation signal---}We now explore the  claim that Model 1 ``is consistent with gravitational dynamics of the dense SYK Hamiltonian beyond its training data''. 
Specifically, the authors claim that the teleportation signal of Model 1 demonstrates behavior compatible with a qubit emerging from a traversable wormhole.
These claims are based upon analyzing the teleportation signal for the pair of operators, $\psi^1$ and $\psi^2$, that were involved in the machine-learning training (Fig.~2 of \cite{jafferis2022traversable}).
To further test whether  Model 1 is consistent with the gravitational dynamics of the dense SYK model, we examine the teleportation signal for operators that were not involved in the training procedure.

Fig.~2 of~\cite{jafferis2022traversable} presents the mutual information of the teleportation protocol as a function of two times: the injection time, $t_0$, and the readout time, $t_1$ (see e.g.~the circuit in Fig.~\ref{fig:teleport}a).  
For the symmetric teleportation protocol, with $t=t_0 = t_1$, traversable wormhole dynamics lead to the presence of a single peak in the mutual information as a function of time~\cite{jafferis2022traversable}. 
We note that Model 1 is trained to reproduce the  symmetric teleportation signal of the SYK model for a specific pair of input Majorana operators, $\psi^1$ and $\psi^2$.
As depicted in Fig.~2a in \cite{jafferis2022traversable} and in Fig.~\ref{fig:teleport}b above, the mutual information indeed exhibits a single peak for the trained operators in Model 1 and for various instances of the SYK model. 
However, for a generic pair of untrained operators in Model 1, the mutual information does not exhibit single-peak behavior, but rather, displays large oscillations as a
function of time that strongly vary for different input operators (Fig.~\ref{fig:teleport}c).
This sharply contrasts with teleportation in the SYK model, where the mutual information exhibits a single consistent peak in time for any pair of input operators (Fig.~\ref{fig:teleport}d). 

We note that Ref.~\cite{jafferis2022traversable} also examines the teleportation protocol at a fixed injection time, while varying the readout time (Fig.~2b in \cite{jafferis2022traversable}); our analysis of this protocol is shown in Appendix \ref{app:fixed_injection}. 
At short times, for Model 1, the mutual information  exhibits a single peak for all pairs of operators, albeit with large variations in peak height.
At longer times, generic pairs of operators exhibit multiple peaks in the mutual information, which are not observed in the SYK model.

\textbf{Size winding---}Beyond a comparison to the $N=10$ SYK model, the authors claim that Model 1 satisfies general behavior predicted by gravity. 
They focus on the property of  size winding (Fig.~3d and~S14 in \cite{jafferis2022traversable}), which is
defined by decomposing a time-evolved operator as follows:
\begin{equation}
\rho_\beta^{1/2} \psi^i(t) = \sum_P c_P \psi^P
\end{equation}
where $c_P$ is a complex coefficient, and $\psi^P$ is a Majorana string with support  $P \in \{ 0, 1 \}^N$.
Here, $P_i = 1$ indicates that $\psi^P$ has support on site $i$.
Size winding is the condition that the phases of the squared coefficients, $c_P^2$, depend linearly on the size $|P|$ of the Majorana string, i.e.~$c_P = e^{i (\alpha |P| / N + \phi) } r_P$ for some real values, $ \alpha,\phi, r_P $.
As in~\cite{jafferis2022traversable}, it is convenient to separate size winding into two distinct properties: 
\begin{enumerate}

\item \emph{Phase alignment}---The phase of $c_P^2$ is equal for all $P$ of the same size. 

\item \emph{Linear slope}---The phase of the sum of coefficients of size $l$, $q(l) = \sum_{|P| = l} c_P^2$, follows a linear slope with respect to $l$.

\end{enumerate}
Analytic calculations show that the SYK model exhibits both properties in the limit of large system sizes~\cite{nezami2021quantum}. 
The authors emphasize that perfect size winding is a necessary criteria of general holographic systems~\cite{size-winding}.

We begin by noting a difference in the structure of time-evolved operators between fully-commuting models such as Model 1, and non-commuting models, such as the SYK model.
In fully-commuting Majorana Hamiltonians, a time-evolved operator has non-zero coefficients for up to $2^{\lfloor N /2\rfloor}$ different strings.
In a non-commuting Majorana Hamiltonian,  a time-evolved operator has  non-zero coefficients for up to $2^{N-1}$ strings.
This difference is particularly pronounced at small system sizes: In Model 1 there are $8$ non-zero coefficients (Fig.~\ref{fig:size-winding}a), while in the $N=7$ SYK model there are $64$ non-zero coefficients  and in the $N=10$ SYK model studied in \cite{jafferis2022traversable} there are $512$ non-zero coefficients.

\emph{Phase alignment---}Ref.~\cite{jafferis2022traversable} quantifies  the degree of phase alignment by considering the ratio:
\begin{equation}
r_l =  \frac{ \left| \sum_{|P| = l} c_P^2 \right| }{\sum_{|P| = l} |c_P|^2},
\end{equation}
which is unity when phase  alignment is perfect.
Using $\psi^1$ (which is one of the operators that was trained upon), the authors show that Model 1 exhibits $r_l \gtrsim 0.95$ for 
all $l$ (Fig.~S14 in \cite{jafferis2022traversable}).
The authors refer to this as perfect size winding. 
In comparison, Ref.~\cite{jafferis2022traversable} finds that the $N=10$ SYK model exhibits  $r_l \sim 0.75$ (Fig.~S19 in \cite{jafferis2022traversable}), which the authors refer to as ``damped'' size winding.

To avoid characterizing the phase alignment ratio for each $l$, we consider the average ratio, $r = \sum_l \left| q(l) \right|$.
Since $r$ is lower bounded by a two-point correlation function, $W = \textrm{tr}( \psi^i  \rho_\beta^{1/2} \psi^i \rho_\beta^{1/2} ) = \sum_P c_P^2$, it is natural to define a rescaled phase alignment ratio as $\bar{r} = \frac{r - W}{1 - W}$ (see Appendix \ref{app:sizewinding} for details).

In Fig.~\ref{fig:size-winding}b, $\bar{r}$ is plotted for Model 1 and the SYK model, for all operators. 
Consistent with the authors' observations (Figs.~S14, S19 in \cite{jafferis2022traversable}), Model 1 exhibits a significantly higher  $\bar{r}$ than the  SYK model.

We begin by examining to what extent the large phase alignment, $\bar{r}$, is a result of the authors' machine learning procedure. 
In particular, we consider Model 1 with \emph{random} numerical coefficients in front of the five Hamiltonian terms.
In all cases, we find that the phase alignment, $\bar{r}$, is similar to Model 1 and significantly higher than the SYK model. 
Three specific instances are shown in Fig.~\ref{fig:size-winding}b. 
Next, we further randomize Model 1, by considering Hamiltonians with five  random, commuting, four-body Majorana terms (in addition to random coefficients).
Again, we find that the phase alignment, $\bar{r}$, is similar to Model 1 and significantly higher than the SYK model (Fig.~\ref{fig:size-winding}b).

The above observations suggest that a large phase alignment, $\bar{r}$, is a generic feature of many fully-commuting models at small system sizes. 
As a further example, we plot the phase alignment, $\bar{r}$, for a  random all-to-all Ising model with four spins, and find similar behavior to Model 1; as shown in Appendix \ref{app:scaling}, we find that the phase alignment, $\bar{r}$, decreases with increasing system size.

\emph{Linear slope---}The authors claim that Model 1 exhibits a linear size-winding slope.
To demonstrate this, in Fig.~3d of \cite{jafferis2022traversable}, the authors plot the phase of $q(l)$ as a function of $l$ for the Majorana operator $\psi^1$ (which was trained upon).
The phase indeed exhibits a linear slope with respect to $l$.

We examine how generic this behavior is, and the extent to which it results from the authors’ machine learning procedure.
In order to better compare across models, we introduce a metric, $\chi$, for quantifying the linear slope property (see Appendix \ref{app:sizewinding} for details).
As depicted in Fig.~\ref{fig:size-winding}c, $\psi^1$ is indeed characterized by a large value of $\chi$.
For the other operator that was trained upon, $\psi^2$, we also find that $\chi$ is large. 
However, for the untrained operators, some Majoranas exhibit large values of $\chi$, while others exhibit small values of $\chi$ implying that they do not satisfy the linear slope property.
This behavior is illustrated in Fig.~\ref{fig:size-winding}d and  contrasts with the SYK model.
For the SYK model, all operators exhibit the linear slope property and $\chi$ is clustered near unity, as shown in Fig.~\ref{fig:size-winding}c.

As in the preceding discussion about phase alignment, the fact that some operators in Model 1 exhibit a linear slope  with large $\chi$, is a generic feature of small-size, fully-commuting  Hamiltonians. 
The distribution of $\chi$ across all operators is illustrated in Fig.~\ref{fig:size-winding}c, for the three types of models considered in the preceding subsection:
(i) Model 1 with random coefficients, (ii) Model 1 with random terms and coefficients, and (iii)  fully-commuting random Ising models.
In each case, the behavior is comparable to Model 1---certain operators exhibit the linear slope property with large $\chi$, while others do not.

We note that for all three models studied by the authors (Model 1 in the main text and Models 2,3 in the Supplemental Material), the trained operators exhibit a relatively high degree of linearity compared to other operators.
This suggests that the authors' machine-learning procedure may have introduced a bias among the trained operators. 
Nevertheless, the distribution of $\chi$ over all operators resembles that of generic random small-size fully-commuting models. 

\begin{figure*}[t]
\includegraphics[width=\textwidth]{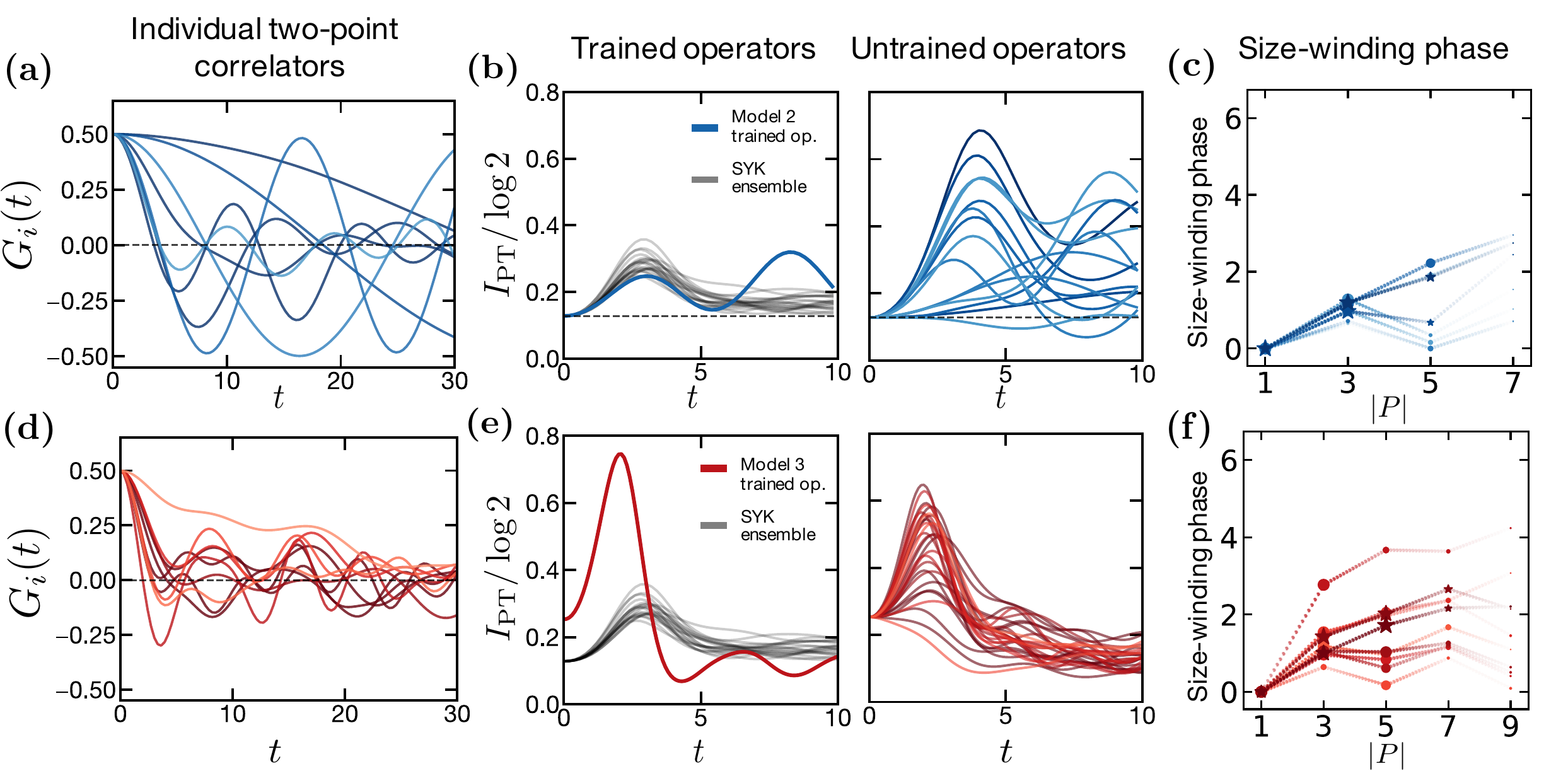}
\caption{%
\textbf{(a)} Two-point correlation functions, $G_i(t)$, for all Majorana operators in Model 2. 
\textbf{(b)} (left) Mutual information for the symmetric teleportation protocol with the trained operators, $\psi^1$ and $\psi^2$, and $\mu = -12$ for: Model 2 (blue), and multiple instances of $N=10$ SYK model (grey). 
(right) Mutual information for the symmetric teleportation protocol with all pairs of untrained operators, and $\mu = -12$. 
\textbf{(c)} Size-winding phase for each of the untrained operators at $t=2.8$. 
\textbf{(d-f)} Depicts the analogous results for Model 3. As in \cite{jafferis2022traversable}, the teleportation protocol  is performed with $\mu=-17$ for Model 3 (replicating Fig.~S25 of \cite{jafferis2022traversable}),  and the size-winding phase is evaluated at $t=2$.}
\label{fig:model2and3}
\end{figure*}

Size winding has recently emerged as a prominent feature of the holographic correspondence for systems with a nearly AdS$_2$ bulk~\cite{brown2019quantum}, leading to speculation that the presence of perfect size winding could be a strong signature of gravity~\cite{nezami2021quantum,schuster2022many}.
The fact that the perfect size winding  observed in~\cite{jafferis2022traversable} seems reliant on small-size, fully-commuting models---which defy other features of holography such as thermalization~\cite{hawking1975particle,horowitz2000quasinormal}, complexity~\cite{susskind2016computational}, and chaos~\cite{maldacena2016bound}---raises the question of whether the observed perfect size winding is indeed connected to gravitational physics in a substantive manner.

\bibliography{refs.bib}

\appendix

\section{Other learned models} \label{app:model2and3}

In the Supplemental Material of \cite{jafferis2022traversable}, two additional learned Hamiltonians are studied numerically. 

\emph{Model 2}---The first of these, which we refer to as Model 2, is  given in Eq.~(S16) of \cite{jafferis2022traversable}:
\begin{equation} \label{eq:model2}
\begin{split}
H = & -0.35 \psi^1 \psi^2 \psi^3 \psi^6 +0.11 \psi^1 \psi^2 \psi^3 \psi^8 -0.17 \psi^1 \psi^2 \psi^4 \psi^7 \\
& -0.67 \psi^1 \psi^3 \psi^5 \psi^7+0.38 \psi^2 \psi^3 \psi^6 \psi^7 - 0.05 \psi^2 \psi^5 \psi^6 \psi^7.
\end{split}
\end{equation}
Model 2 is produced from the same machine-learning procedure as Model 1, i.e.~designed to match the teleportation signal of the $N=10$ SYK model.
The authors claim that Model 2 demonstrates perfect size winding  and ``is consistent with other gravitational signatures''.

As noted in \cite{jafferis2022traversable}, Model 2 is not fully commuting. 
Nevertheless, we observe that Model 2 becomes fully-commuting if: (i) the two \emph{smallest} terms in Eq.~\eqref{eq:model2} are removed, and (ii) one performs a basis rotation:
\begin{equation}
\begin{split}
\psi^1 & \rightarrow \cos(\theta) \psi^1 + \sin(\theta) \psi^7, \\
\psi^7 & \rightarrow \cos(\theta) \psi^7 - \sin(\theta) \psi^1,
\end{split}
\end{equation}
with $\theta = \tan^{-1}(-0.35/0.38)$. 
At the timescale of teleportation ($t = 2.8$), the two smallest terms provide relatively small corrections to physical observables.
Thus, Model 2 can be considered weakly perturbed from a fully-commuting limit.

Consistent with this observation, we find that our main observations regarding Model 1 also apply to Model 2 (Fig.~\ref{fig:model2and3}a-d).
In particular, the individual two-point correlation functions exhibit strong revivals, the teleportation signal does not resemble the SYK model for untrained operators, and the size winding behavior resembles that of a random fully-commuting Hamiltonian (Fig.~\ref{fig:size-winding}c).
In addition, we note that the teleportation signal  for the \emph{trained} operators in Model 2 displays a significant revival within the timescale on which it was trained (Fig.~\ref{fig:model2and3}b).
This contrasts with the $N=10$ SYK model and indicates that the training procedure was not fully successful; such disagreement is not shown or commented on in \cite{jafferis2022traversable}. 

\emph{Model 3}---The second additional model, which we refer to as Model 3, is given in Eq.~(S17) of~\cite{jafferis2022traversable}:
\begin{equation} \label{eq:model3}
\begin{split}
H &= 0.60 \psi^1 \psi^3 \psi^4 \psi^5 +0.72 \psi^1 \psi^3 \psi^5 \psi^6 +0.49 \psi^1 \psi^5 \psi^6 \psi^9 \\
& +0.49 \psi^1 \psi^5 \psi^7 \psi^8+0.64 \psi^2 \psi^4 \psi^8 \psi^{10} - 0.75 \psi^2 \psi^5 \psi^7 \psi^8 \\
& +0.58 \psi^2 \psi^5 \psi^7 \psi^{10} - 0.53 \psi^2 \psi^7 \psi^8 \psi^{10}.
\end{split}
\end{equation}
Model 3 is produced via a different machine-learning procedure, which is designed to optimize the asymmetry in the teleportation signal between positive and negative couplings.
Unlike Models 1 and 2, Model 3 is not fully-commuting or near fully-commuting.

Referring to the average two-point correlator, the authors demonstrate that ``no periodicities are present despite the small number of terms in the Hamiltonian'' (Fig.~S26 of~\cite{jafferis2022traversable}).
In Fig.~\ref{fig:model2and3}d, we observe that the individual two-point correlators also exhibit thermalizing behavior at long time scales ($t \sim 30$).
This is consistent with Model 3 being non-commuting.
At earlier times, the correlators exhibit oscillations that are smaller than those of Model 1 and 2, but larger than fluctuations in the $N=10$ SYK model.

The teleportation signal for Model 3 exhibits a single-peak structure for nearly all operators, albeit with large variations in peak height (Fig.~\ref{fig:model2and3}e).

The authors note that Model 3 does not exhibit perfect size winding, but rather features a ``consistently large ratio [of phase alignment], suggesting slightly damped size winding''.
Indeed, we find that the phase alignment, $\bar r$, for Model 3 is comparable to that of the $N=10$ SYK model and lower than that of small-size fully-commuting models (Fig.~\ref{fig:size-winding}b).
This is consistent with our observation that perfect phase alignment at small system sizes is a generic feature of fully-commuting Hamiltonians and not of non-commuting Hamiltonians. 
We note that only some operators in Model 3 (including the trained operators) exhibit a high degree of linearity $\chi$ (Fig.~\ref{fig:size-winding}c).
In this respect, Model 3 resembles the behavior of fully-commuting or nearly fully-commuting models (including Model 1 and 2) and not the SYK model.

\begin{figure}
\includegraphics[width=0.8\linewidth]{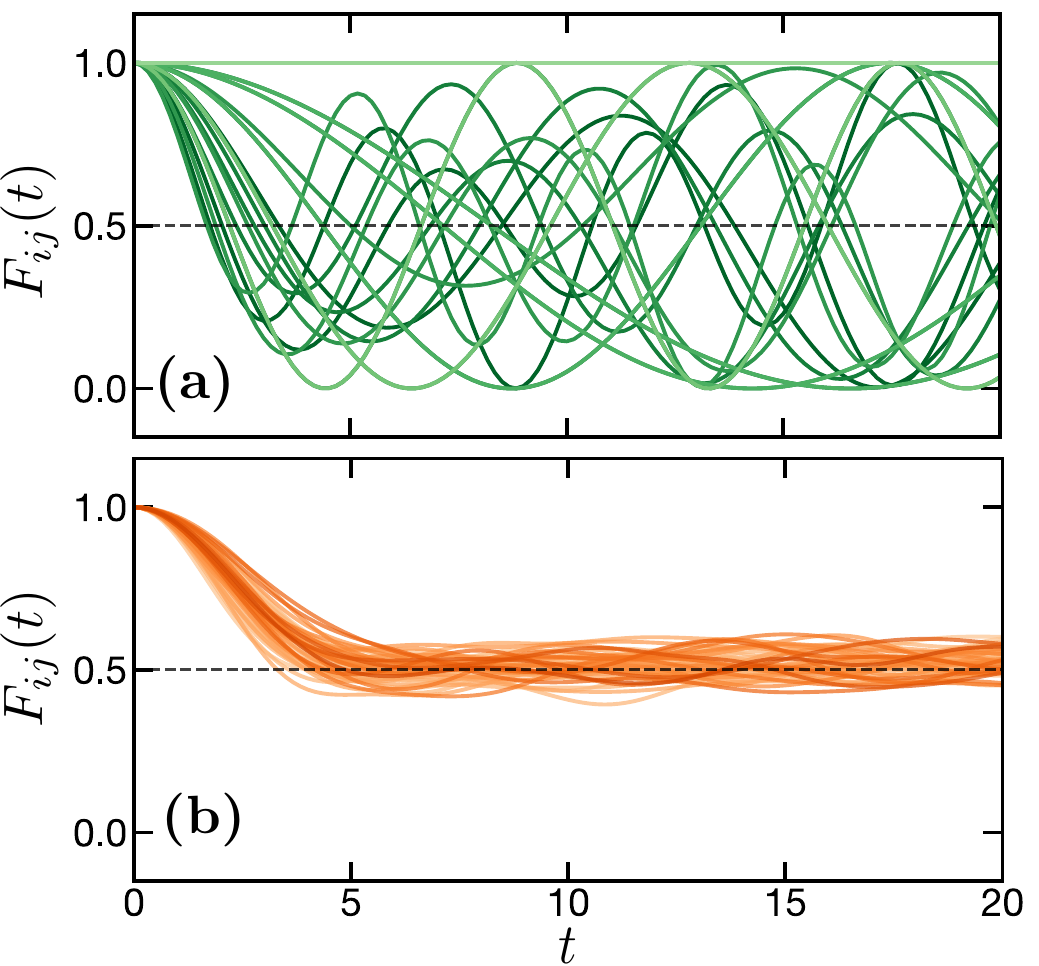}
\caption{%
\textbf{(a)} Four-point correlation functions, $F_{ij}(t)$, for Model 1, shown for all pairs of Majorana operators, $i<j\in [1,7]$. \textbf{(b)} The same correlation functions for a specific instance of the $N=10$ SYK model with $J=1.125$ and $i<j\in [1,10]$.
}
\label{fig:4pt}
\end{figure}

\section{Four-point correlators with $i \neq j$}

Scrambling is quantified in~\cite{jafferis2022traversable} via the behavior of the  four-point correlation functions, $F_{\textrm{avg}}(t) = \sum_{i=1}^8 F_i(t)$, with $F_i (t) =- \textrm{Re} \left [ \big\langle \left [\psi^i(t), \psi^i(0) \right ]^2 \big\rangle_\beta \right ]$.
We note that such correlation functions, consisting of the same Majorana $\psi^i$ for the time-evolved and static operators, are not the most direct probe of scrambling dynamics, since their initial growth occurs on the same timescale as the decay of two-point correlation functions (i.e.~the thermalization time).
In the SYK model at large system sizes, the initial growth reaches a value of unity and is followed by a slower decay to value $1/2$ on the timescale of the scrambling time; such non-monotonic behavior is evident in the time traces of the $N=10$ SYK model shown in Fig.~\ref{fig:2pt}d.
A more typical probe of scrambling is the four-point correlator, $F_{ij} (t) =- \textrm{Re} \big[ \big\langle \left [\psi^i(t), \psi^j(0) \right ]^2 \big\rangle_\beta \big]$, for different operators, $i \neq j$.
In the SYK model at large system sizes, this correlator decays monotonically from unity to value $1/2$ on the timescale of the scrambling time.

\begin{figure*}
\includegraphics[width=\textwidth]{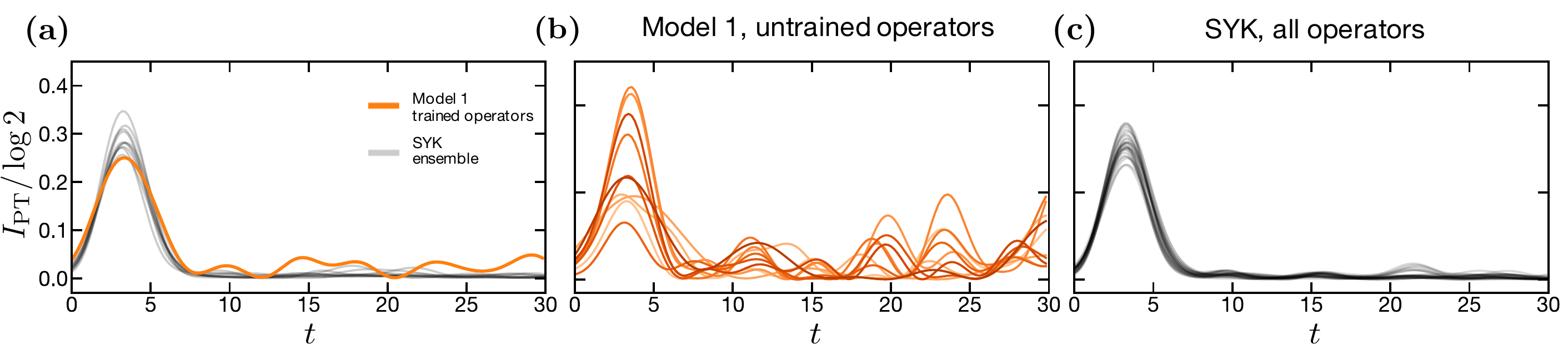}
\caption{%
\textbf{(a)} Mutual information of the teleportation protocol with fixed injection time as a function of the readout time (i.e.~$t_0 = 2.8$ and $t = t_1$). The mutual information for Model 1 and the trained operators, $\psi^1$ and $\psi^2$, is in reasonable agreement with that of multiple instances of the $N=10$ SYK model (grey). \textbf{(b)} The mutual information for Model 1 and all pairs of untrained operators, $\psi^i$ and $\psi^j$ with $i<j\in [2,7]$, exhibits variations and revivals as a function of time. \textbf{(c)} The mutual information for all pairs of operators in the $N=10$ SYK model exhibits a single consistent peak.}
\label{fig:teleport-fixed}
\end{figure*}

In Fig.~\ref{fig:4pt}, we plot the four-point correlation functions, $F_{ij} (t)$ with $i\neq j$, for both Model 1 and the $N=10$ SYK model.
Much like the four-point correlation functions with $i = j$ (i.e.~$F_i(t)$, see Fig.~\ref{fig:2pt}), we find that the four-point correlation functions in Model 1 exhibit strong oscillations in time for all $i \neq j$.
In fact, the oscillations for many pairs of operators have unit amplitude. 
In contrast, in the $N=10$ SYK model, all correlation functions exhibit a smooth decay to value $1/2$.

\section{Teleportation at fixed injection time}\label{app:fixed_injection}

As previously discussed, two versions of the teleportation protocol are analyzed in \cite{jafferis2022traversable}: using symmetric injection / readout times and fixed injection time. 
In Fig.~\ref{fig:teleport-fixed}, we present results for latter protocol for Model 1 and the $N=10$ SYK model.
For Model 1, when the protocol is performed with the pair of operators that were trained on, the mutual information displays a single peak as a function of time. 
For other pairs of operators, the mutual information displays an initial peak, whose height varies significantly for different pairs of operators, followed by revivals at later times.
This contrasts with the SYK model, in which the mutual information displays a single consistent peak for all pairs of operators, with small and infrequent fluctuations at late times.

\section{Size-winding metrics} \label{app:sizewinding}

Here, we elaborate on the phase alignment, $\bar r$, and the linear slope metric, $\chi$, which are plotted in Fig.~\ref{fig:size-winding}b and Fig.~\ref{fig:size-winding}c.

\emph{Phase alignment}---We recall that in~\cite{jafferis2022traversable}, the phase alignment is quantified by plotting the ratio, $r_l =  \left| \sum_{|P| = l} c_P^2 \right| / \sum_{|P| = l} |c_P|^2$, for different sizes $l$ (Figs.~S14 and~S19 of~\cite{jafferis2022traversable}).
The denominator of this quantity is the operator size distribution, $p(l) = \sum_{|P|=l} |c_P|^2$, which is normalized to one, $\sum_l p(l) = 1$.
To facilitate comparison between different operators and models, we consider the weighted average of $r_l$, $r = \sum_l p(l) \, r_l = \sum_l \left| \sum_{|P| = l} c_P^2 \right| = \sum_l \left| q(l) \right|$.
For a given Hamiltonian, $r$ is lower bounded by the two-point function, $W = \tr( \psi^i(t) \rho^{1/2} \psi^i(t) \rho^{1/2} ) = \sum_P c_P^2$.
We note that this two-point function is constant in time, and therefore the sum of the squared coefficients is also constant in time.
Taking into account this lower bound motivates us to rescale $r$ as $\bar r = \frac{r - W}{1 - W}$, which ranges from zero to one.

\emph{Linear slope}---We seek to quantify the degree to which the phases of $q(l)$ follow a linear slope with respect to the size $l$.
The fit of a line of slope $\mu$ can be quantified via $C(\mu) = \left| \sum_l q(l) e^{- i \mu l} \right|$.
When deviations from a linear slope are small, this reduces to unity minus a weighted sum of squared errors; when deviations are large, it takes into account the periodicity of the phases. 
The best fit, $C^*$, is found by maximizing over $\mu$, $C^* = \max_\mu C(\mu)$.

We define the metric, $\chi$, to interpolate between zero and one as $C^*$ interpolates between its minimum and maximum values.
The maximum value of $C(\mu)$ is given by the weighted average, $r$, of the phase alignment ratio.
The minimum value is lower bounded by the two-point function, $W = C(\mu = 0)$.
In addition, at small system sizes it is relevant to consider a second lower bound, corresponding to fitting a line between the two coefficients, $q(l_1)$ and $q(l_2)$, with the largest magnitude.
This consideration is necessary to avoid concluding that functions $q(l)$ with support on only two values of $l$ have non-trivial size winding.
This fit produces a $C$ of value at least $M = | q(l_1) | + | q(l_2) | - (r-| q(l_1) | + | q(l_2) |) = 2 | q(l_1) | + 2| q(l_2) |-r$.
We thus define the metric,
\begin{equation}
    \chi = \frac{C^* - L}{r-L},
\end{equation}
where $L$ is the larger of the two lower bounds, $L = \max(W,M)$.

\section{Other fully-commuting models} \label{app:scaling}

Here we include details on the random fully-commuting models presented in Fig.~\ref{fig:size-winding}. For all random models in Fig.~\ref{fig:size-winding}, we take $\beta = 4$ and $t = 2.8$, identical to Model 1.

\emph{Majorana models}---In Model 1 with randomized coefficients, we draw each coefficient from a normal distribution with mean zero and standard deviation equal to the root-mean-square of the coefficients of Model 1. 
In Model 1 with randomized terms and coefficients, we generate five random fully-commuting terms by successively drawing random four-Majorana terms (from $N=7$ total Majorana operators) and keeping each term only if it commutes with all terms already kept.

\emph{Ising models}---We consider random all-to-all Ising models with Hamiltonian, $H = \frac{1}{\sqrt{N}} \sum_{i < j} J_{ij} Z^i Z^j$.
The coefficients $J_{ij}$ are drawn from a normal distribution with mean zero and standard deviation $J = 0.17$.

\begin{figure}
\includegraphics[width=0.85\columnwidth]{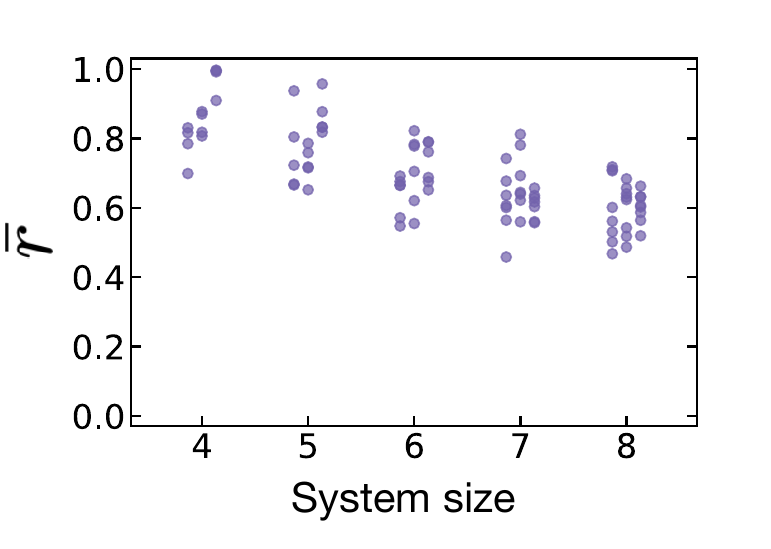}
\caption{Phase alignment, $\bar r$, of the random all-to-all Ising model as a function of the system size $N \in [4,8]$, with $J = 0.17$ and $\beta = 4$. 
Three disorder realizations are shown at each system size, with small horizontal offsets for clarity.
}
\label{fig:scaling}
\end{figure}

\emph{Finite-size scaling}---To explore whether the size winding behavior of fully-commuting models persists at larger system sizes, in Fig.~\ref{fig:scaling} we plot the phase alignment, $\bar r$, for random all-to-all Ising models as a function of the system size $N$.
We focus on Ising models to avoid subtleties with scaling random fully-commuting Majorana models to larger system sizes (namely, there is no canonical choice of which fully-commuting terms to include). 
We scale the evolution time $t$ with the square root of the system size, $t = 2.8 \sqrt{N / 4}$, to ensure that operators grow to the same fraction of the system size for each $N$.
We find that the phase alignment, $\bar r$, exhibits a decreasing trend with the system size.

\section{Missing parameters}
Several parameters are omitted from \cite{jafferis2022traversable} which are necessary for reproducing the numerical results shown.
For the ease of future studies, we list these parameters below:
\begin{itemize}
\item In plots showing mutual information (e.g.~Figs.~1a, 3a, 3e of \cite{jafferis2022traversable}), the mutual information is divided by $\log(2)$.
\item For the teleportation plots (e.g.~Figs.~1a, 3a, 3e of \cite{jafferis2022traversable}), the two-sided coupling is $V = \frac{1}{qN} \sum_i \psi_L^i \psi_R^i$ with $N=10$ and $q=4$, for both Model 1 and the SYK models. The teleportation protocol is performed using $\psi^1$ and $\psi^2$.
\item In Fig.~3f of \cite{jafferis2022traversable}, the instantaneous coupling plot uses $\mu = -12$, however with $V$ normalized using $N=8$. This is distinct from the other plots in Figs.~1-3, which normalize $V$ with $N=10$. The Trotterized coupling plot uses $\mu = -18$ and $N=8$, and is Trotterized into three time steps, $t = -1.6,0,1.6$.
\item As previously emphasized, the correlation functions in Fig.~3b of \cite{jafferis2022traversable} are averaged over all local Majorana operators (for Model 1, this consists of $8$ operators: the 7 operators in Eq.~(\ref{eq3}) and an additional uncoupled operator). Also, the authors plot the real part of two- and four-point correlation functions. 
To summarize, the plots correspond to ~$G_\textrm{avg}(t) = \frac 1 {8} \sum_{i=1}^8 G_i(t)$ with $G_i(t) =  \textrm{Re}[ \left<\psi^i(t) \psi^i(0)  \right>_\beta ]$ and $F_\textrm{avg} (t)   = \sum_{i=1}^8 F_i (t)$ with $F_i (t) =- \textrm{Re} [ \big\langle \left [\psi^i(t), \psi^i(0) \right ]^2 \big\rangle_\beta ]$. 
\item We are not able to exactly replicate the SYK dynamics shown in Figs.~1 and 3b in \cite{jafferis2022traversable}. We find qualitatively good agreement using an ensemble of $N=10$ SYK models with $J = 1.25$ for Fig.~1 in \cite{jafferis2022traversable} and $J = 1.125$ for Fig.~3b in \cite{jafferis2022traversable}.
\end{itemize}

\end{document}